\documentclass[12pt]{JHEP}
\usepackage{amsmath,amssymb,epsfig}

\newcommand{\be}{\begin{equation}}
\newcommand{\ee}{\end{equation}}
\newcommand{\beqa}{\begin{eqnarray}}
\newcommand{\eeqa}{\end{eqnarray}}

\def\eeq{\end{equation}}

\newcommand{\HHH}{{\cal H}}

\relax

\title{Jet quenching at finite `t Hooft coupling and chemical
potential from AdS/CFT}

\author{N\'estor Armesto${}^{\,*}$, Jos\'e D. Edelstein${}^{\,*\dagger}$ and
Javier Mas${}^{\,*}$\\
\vspace{0.1in}

${}^{\,*}$Departamento de F\'\i sica de Part\'\i culas and IGFAE,
Universidade de Santiago de Compostela, E-15782 Santiago de Compostela, Spain
\vspace{0.1in}

${}^{\,\dagger}$Centro de Estudios Cient\'\i ficos (CECS), Casilla 1469,
Valdivia, Chile
\vspace{0.1in}

E-mail addresses: {\tt nestor,edels,jamas@fpaxp1.usc.es}
}

\abstract{Following the nonperturbative prescription for the jet quenching
parameter recently proposed by Liu, Rajagopal and Wiedemann, we compute the
first correction in the inverse `t Hooft coupling corresponding to string
$\alpha'$ corrections in the dual background. We also consider the
introduction of a chemical potential for the $U(1)^3$ gauged R-symmetry.
While the former mildly diminishes the jet quenching parameter --this
suggesting a smooth interpolation between the strong coupling and
perturbative results--, the latter generically increases its value. We comment on the 
extension of this setup to quarks of finite mass. }

\preprint{CECS-PHY-06/17 \\ 22 June 2006 \\ hep-th/0606245}

\begin{document}


\section{Introduction and Summary}

The experimental program at the Relativistic Heavy Ion Collider (RHIC) in
Brookhaven National Laboratory \cite{RHIC}, has provided considerable
insight into the properties of strongly interacting matter at high energy densities. Phenomenological analysis has established several striking
features of such substance. First, the results on elliptic flow are well
described by hydrodynamical models only if the shear viscosity is
taken very low. The medium behaves like a strongly coupled plasma which resembles a liquid more than the gas of quasi-free partons long expected
to be the state of matter at such energy densities, known as the Quark Gluon Plasma (QGP). As a second indication of this behavior, high energy partons
traversing the medium are strongly quenched. This phenomenon is usually characterized, in models of medium-induced radiation, by the so-called
quenching parameter (or transport coefficient) $\hat{q}$
\cite{Baier:1996sk}. This parameter has the meaning of the average
squared transverse momentum transferred from the medium to the traversing
parton, per unit mean free path
(see the reviews \cite{Reviews,Kovner:2003zj}).

Phenomenological models differ in the detailed framework for calculating
the radiative energy loss \cite{Baier:1996sk,RadEnLoss}, in the treatment of
the geometry and dilution of the medium \cite{Medium}, as well as in the
consideration of flow-induced radiation \cite{FIR} and of additional elastic
scattering \cite{Adil:2006ei}. The extracted values of the transport
coefficient are $\hat{q}\sim 1\div 15$ GeV$^2$/fm, substantially larger
than those found in studies of hadron production in DIS on nuclear targets,
see e.g. \cite{Arleo:2003jz}. While the lower bound is compatible with expectations from perturbative QCD \cite{Baier:2002tc}, higher values
demand additional non-perturbative mechanisms. Therefore, and while waiting
for upcoming both experimental and phenomenological efforts, it is of
uttermost importance to get further information on the possible values
of $\hat{q}$ in the strong coupling limit.

The traditional tool for such studies, namely lattice QCD, cannot be
presently applied to determining the jet quenching parameter. In contrast,
AdS/CFT duality \cite{Maldacena:1997re} provides a powerful calculational framework where quantum properties of supersymmetric Yang--Mills theories
at strong 't Hooft coupling $\lambda=g_{YM}^2 N_c$ and large
number of colors $N_c$, are translated into classical computations in a gravitational background. The applications of AdS/CFT techniques to thermal deformations of gauge theories started in \cite{Witten:1998zw}. There, the thermodynamics of the black brane geometry was conjectured to describe the behavior of the dual quantum field theory at the Hawking temperature of the black hole. In the limit of a flat horizon, the dual field theory lives in
an unconfined phase at strong coupling. Placing probe charges in such a background corresponds to the insertion of sources for very massive quarks
in the dual theory. Following this lore, a number of important results where derived concerning the $q\bar q$ potential, including features like confinement and screening both at zero and finite temperature \cite{Rey:1998ik}. They typically involve a Wilson line stretching either in a timelike or a spacelike direction; rotating Wilson lines where also examined as
putative duals to high spin mesons. More recently, the study of
thermodynamical properties was extended to encompass near equilibrium
magnitudes \cite{Policastro}. A not minor surprise came out with the
finding of a universal ratio between the shear viscosity and the entropy
density, $\eta/s = 1/(4\pi)$ \cite{UnivShear} for quantum field theories
admitting a holographic dual description. This ratio was conjectured to
set a universal bound on physical thermal field theories. The data at
RHIC suggest that the values for the QGP are compatible with the lower
bound, this strongly supporting the use of AdS/CFT to describe such a
system.

Motivated by these successful applications of AdS/CFT to the study of
strongly coupled phenomena in thermal gauge theories, Liu, Rajagopal and Wiedemann (LRW) recently proposed a scheme to determine the jet quenching parameter
\cite{Liu:2006ug}\footnote{After the initial proposal, a host of papers
have appeared \cite{Buchel:2006bv,HatQ,Caceres:2006as,Lin:2006au,Avramis:2006ip} which have shown that this result
is not universal. Also the very interesting and related problem of a drag
force on the brane has been addressed by several authors \cite{Drag}.}.
In their construction, it is central to use the identification \cite{Kovner:2003zj} of this parameter with the coefficient in the exponent of an adjoint Wilson loop computed along a rectangular contour $\mathcal{C}$ with a large
distance $L^-$ along the light-cone, and a spacelike separation $L$,
~$L\ll L^-$:
\beqa
\langle W^A(\mathcal{C})\rangle \equiv \exp\left[-\frac{1}{4} \hat{q} L^- L^2
\right] ~.
\label{intro-eq1}
\eeqa
At large $N_c$ this Wilson loop can be expressed in terms of the Wilson
loop for the fundamental represention, $\langle W^A(\mathcal{C})\rangle
\simeq \langle W^F(\mathcal{C})\rangle^2$. In turn, the AdS/CFT
correspondence tells us that this fundamental Wilson loop can be computed
\cite{Rey:1998ik} evaluating the classical Nambu-Goto action $S$ for
a string ending on the boundary along the previous contour,
\beqa
\langle W^F(\mathcal{C})\rangle = \exp\,[-S(\mathcal{C})] ~.
\label{intro-eq2}
\eeqa
The result in \cite{Liu:2006ug}, obtained in a near extremal D3 background 
corresponding to ${\cal N}=4$ SUSY QCD at finite temperature, exhibited some interesting features.\footnote{Although the relation between the results
in this framework and real QCD is unclear, the hope is that an understanding
of the size and dependences of the jet quenching parameter could provide
some kind of upper bound, while the perturbative QCD results should provide
a lower bound. In this way, computing the results in less supersymmetric backgrounds might provide an indication (assuming a smooth behavior for the transition from ${\cal N}=4$ SUSY QCD to real QCD) of the expected value for this parameter at strong coupling.} The quenching parameter $\hat{q}$
turned out to be proportional to $T^3$ (which of course provides the correct dimensions) and to $\sqrt{\lambda}$ (thus to $\sqrt{N_c}$), the latter being totally different to the {\it a priori} expected dependence on the number of degrees of freedom of the energy or entropy densities, hence $\propto N_c^2$.
In this way, the quenching parameter appears not to be a direct measure of the
energy density of the system, $\hat q \propto \epsilon^{3/4}$ as usually
assumed \cite{Baier:2002tc}, but of the third power of the temperature.
Moreover, the numerical values turned out to be astonishingly close to the experimental data: for standard values $\alpha_s = \frac{1}{2}$ and
$N_c=3$, $\hat{q} = 0.94\, (3.16)$ GeV$^2$/fm for $T = 200\, (300)$ MeV.
Notice, however, that these quantities imply a `t Hooft coupling
$\lambda = 6\pi$,
while the gravity computation is strictly valid in the limit $\lambda \to
\infty$. Thus, a more precise theoretical value of $\hat{q}$ in the strong coupling limit neatly demands the understanding of finite `t Hooft coupling corrections to the result in \cite{Liu:2006ug}. This is one of the targets
in this letter.

The paper is organized as follows. In the next section, trying to keep
ourselves as generic as possible, we provide a formula for the jet quenching
parameter that can be readily applied to a large class of metrics. As a quick example, we present the
results for the thermal deformation of Witten's D4--brane background
\cite{Witten:1998zw}. We also comment on the fact that this formula admits a straightforward generalization to encompass quark sources of finite mass, and provide some
preliminar numerical analysis. In section 3 we compute the first correction in
the inverse `t Hooft coupling to the value given by LRW. We show that this correction mildly diminishes the jet
quenching parameter. This suggests a smooth interpolation between the
strong coupling regime and the perturbative results, in analogy with what
has been observed for the free energy and the ratio $\eta/s$
\cite{Gubser:1998nz,Buchel:2004di}. In section 4, the effect of turning on
chemical potentials\footnote{These chemical potentials are conjugated to
$R$--charge densities of $\mathcal{N}=4$ SYM theory. They should not be
confused with that corresponding to the baryon density in QCD whose
implementation in the dual supergravity side is currently an open problem.
Indeed, the baryonic charge is not dual to a $U(1)$ isometry of the supergravity
background. We thank Krishna Rajagopal for stressing the importance
of this point.} is thoroughly investigated. The relevant metric
corresponds to the background of a stack of rotating D3--branes with
maximal number of angular momenta. We explore the evolution of the
jet quenching parameter along the space of these three independent charges.
We typically find an enhancement within the range of thermodynamical
stability. We further compare with recent results which have appeared
on the subject \cite{Caceres:2006as,Lin:2006au,Avramis:2006ip}.
 

\section{The jet quenching parameter}

In this section we shall provide a formula that allows to readily compute
$\hat q$ in  string theory backgrounds within the class of metrics that are suitable for an AdS/CFT duality, including the case in LRW. We will follow essentially the same steps as in \cite{Buchel:2006bv}. The family of ten
dimensional metrics of interest for us adopt the following form:
\beqa
ds^2 &=& G_{MN}\, dX^M dX^N \nonumber\\
&=& - c_T^2\, dt^2 + c_X^2\, dx^i dx_i + c_R^2\, dr^2 + G_{M n} dX^M dX^n ~,
\label{classmet}
\eeqa
where $X^M = (t, x^i, r; X^n), ~i=1,...,p, ~n=1,...,8-p$. This class of
metrics encompasses rotating backgrounds which we shall analyze later. We are interested in black brane solutions. 

Following \cite{Liu:2006ug} we will consider the following lightlike
Wilson line
\be
x^- = \tau ~, \qquad x^2 = \sigma ~, \qquad r = r(\sigma) ~,
\ee
with $x^\pm = (x^0 \pm x^1)/\sqrt{2}$, $\tau\in (0, L^-)$,
$\sigma\in (-\frac{L}{2},\frac{L}{2})$ and
$L^- \gg L$ so that isometry along $x^-$ direction holds approximately. Also
we shall take a symmetric configuration around $\sigma = 0$, hence $r_0
= r(0)$ is an extremal point, $r'(0) = 0$. The induced metric reads as follows
\be
g_{\tau\tau} = \frac{1}{2}( -c_T^2  + c_X^2) ~, \qquad
g_{\sigma\sigma} =  c_X^2 + c_R^2\, r'(\sigma)^2 ~.
\ee
From these expressions, the Nambu--Goto action takes the following form
\be
S = \frac{L^-}{\sqrt 2\pi \alpha'}  \int_{0}^{L/2} d\sigma ~\left( c_X^2
- c_T^2 \right)^{1/2} \left( c_X^2 + c_R^2\, r'(\sigma)^2 \right)^{1/2} ~.
\ee
The energy is a first integral of motion, from which the following equation
for the profile $r(\sigma)$ can be extracted
\be
r'(\sigma)^2 = \frac{c_X^2}{c_R^2} \left( k\, c_X^2\, (c_X^2 - c_T^2 )
- 1 \right) ~,
\label{pend}
\ee
where $k$ is an integration constant.
Let us assume that the r.h.s. of (\ref{pend}) does not vanish at any location
$r\in (r_H,\infty)$ with $r_H$ the location of the horizon, while, on the
other hand, we assume $c_R(r_H) = \infty$. Then, it necessarily holds that
$r_0 = r_H$, and the Wilson line extends symmetrically from $r = \infty$
down to $r_H$. With these assumptions, which we must verify case by case,
the profile can be obtained from
\be
\sigma(r) = \int_{r_H}^r \frac{c_R}{c_X}\, \frac{dr}{\left( k\,
c_X^2\, (c_X^2 - c_T^2) - 1 \right)^{1/2}} ~.
\label{sigmar}
\ee
In particular, the integration constant $k$ is linked with $L$ by
the relation $\sigma(\infty) = \frac{L}{2}$. Going to a dimensionless radial coordinate $\rho = r/r_H$, this is
\be
L  = 2\,r_H\, \int_{1}^\infty \frac{c_R}{c_X}\,
\frac{d\rho}{\left( k\, c_X^2\, (c_X^2 - c_T^2) - 1 \right)^{1/2}} ~.
\ee
The prescription  in LRW for $\hat q$ calls for the leading behavior with
$L$ in the limit $LT \ll 1$. This is clearly related to the limit $k\to
\infty$, i.e.,
\be
L =\frac{2\, r_H}{\sqrt{k}} \int_{1}^\infty \frac{c_R\,
d\rho}{c_X^2\, (c_X^2 - c_T^2)^{1/2}} + {\cal O}(k^{-3/2}) ~.
\label{laele}
\ee
Using (\ref{pend}) and (\ref{sigmar}), we can write the action as follows
\be
S = \frac{r_H\, L^-}{\sqrt{2}\pi \alpha'} \int_{1}^\infty 
\frac{\sqrt{k}\, (c_X^2 - c_T^2)\, c_X\, c_R\, d\rho}{\left(
k\, c_X^2\, (c_X^2 - c_T^2) - 1 \right)^{1/2}} ~.
\ee
We must still subtract the contribution corresponding to the self-energy
of the quarks. This is given by the Nambu--Goto action for a pair of Wilson
lines that stretch straight from the boundary to the horizon,
\be 
S_0 = \frac{r_H\, L^-}{\sqrt 2\pi \alpha'} \int_{1}^\infty c_R\,
(c_X^2 - c_T^2)^{1/2}\, d\rho ~.
\ee
To leading order in $k^{-1}$, taking into account that $L$ is given by
(\ref{laele}), $S_I = S - S_0$ reads \cite{Buchel:2006bv}
\be
S_I = \frac{L^-}{\sqrt{2}\pi \alpha'}\, \frac{L^2}{8 r_H}
\left( \int_{1}^\infty \frac{c_R\, d\rho}{c_X^2\, (c_X^2 -
c_T^2)^{1/2}} \right)^{-1} ~.
\ee
From here, let us extract an expression for the jet quenching parameter.
We find it convenient to define
\be
c_T^2(\rho) =  \frac{1}{\Delta_R }\, \hat c_T^2(\rho) ~, ~~~~~~
c_X^2(\rho) = \frac{1}{\Delta_R }\, \hat c_X^2(\rho) ~, ~~~~~~
c_R^2(\rho) =  \Delta_R\, \hat c_T^2(\rho) ~,
\ee
where the dimensionless quantity $\Delta_R$ reads
\be
\Delta_R = \Bigg( \frac{(\alpha')^{5-p}\, \lambda}{r_H^{7-p}} \Bigg)^{1/2},
\label{deltar}
\ee
$\lambda$ being the `t Hooft coupling in the $p+1$ dimensional dual gauge
theory. From these formulas, (\ref{intro-eq1}) and (\ref{intro-eq2}), we
obtain
\be
\hat q = \frac{1}{\sqrt{2}\pi \lambda }
\left( \frac{r_H}{\alpha'} \right)^{6-p} \,
\left( \int_{1}^\infty \frac{\hat c_R\, d\rho}{\hat c_X^2\, (\hat c_X^2 -
\hat c_T^2)^{1/2}} \right)^{-1} ~.
\label{gjetq}
\ee
As it stands, this formula calls for a translation of $r_H$ in terms of the field theoretical quantities. In the case of non-rotating backgrounds we can provide a more explicit
solution. For this class of metrics
the Hawking temperature   is given by 
the standard formula
\be
T =\left. \frac{1}{4\pi} \frac{ {c_T^2}'(r)}{\sqrt{c_T^2(r)\,
c_R^2(r)}}\right\vert_{r=r_H} ~.
\label{temp}
\ee
Using this definition of the temperature, we can solve for $r_H/\alpha'$ as follows
\be
\frac{r_H}{\alpha'} = \Bigg[ 4\pi \sqrt{\lambda}\, T\, \left( \frac{\sqrt{\hat c_T^2
(\rho)\, \hat c_R^2 (\rho)}}{{\hat c_T^2}{'} (\rho)} \right)\Bigg|_{\rho = 1} \Bigg]^{\frac{2}{5-p}} ,
\label{solverh}
\ee
and replace it in (\ref{gjetq}) to arrive at the formula
\be
\hat q = \frac{1}{\sqrt{2}\pi}\, \Bigg[ 16\pi^2 \Bigg( \frac{\sqrt{\hat c_T^2
(1)\, \hat c_R^2 (1)}}{{\hat c_T^2}{'} (1)} \Bigg)^2\,
\Bigg]^{\frac{6-p}{5-p}}\, T^2\, \left( T^2\, \lambda \right)^{\frac{1}{5-p}}\,
\left( \int_1^{\infty} \frac{\hat c_R \,d\rho}{\hat c_X^2\, (\hat c_X^2 - \hat
c_T^2)^{1/2}} \right)^{-1}.
\label{formulita}
\ee
This expression is   invariant under
reparameterization of the radial coordinate (upon suitable change of the integration limits). Indeed, the dependence on $T$ and $\lambda$ for generic $p$
also coincides with the discussion in LRW. For example, as it stands, it can be directly used to
extract the quenching parameter for the thermal deformation of
Witten's D4--brane background.
In this metric, the fifth dimension has been compactified to a circle
of radius $\ell$. Hence, the four dimensional effective coupling is $\tilde\lambda = \lambda/\ell\equiv  4\pi \alpha_{SYM}N_c$. Therefore we may write for the effective
quenching parameter the following expression
\be
\hat q \simeq 14.26 \,c\, T^3\,\alpha_{SYM}\,N_c ~,
\label{dcuatro}
\ee
where $c=\ell T$ is the ratio of radii of the thermal and Kaluza--Klein
circles. The particular value $c=1$ signals the confinement/deconfinement
transition temperature $T = T_c = 1/\ell$ \cite{Kruczenski:2003uq}.
Therefore, strictly speaking, (\ref{dcuatro}) is valid for $c\geq 1$,
or $T\geq T_c$. The numerical factor in (\ref{dcuatro}) is the
result of (\ref{formulita}) with $p=4$ after inserting $\hat c_T^2
=\sqrt{8\pi}\rho^{3/2}(1-\rho^{-3})$, $\hat c_X^2 =\sqrt{8\pi}\rho^{3/2}$
and $\hat c_R^2 = (8\pi)^{-1/2}\rho^{-3/2}(1-\rho^{-3})^{-1}$
\cite{Itzhaki:1998dd}. For standard values $\alpha_s = \frac{1}{2}$
and $N_c=3$, we get $\hat{q} = 0.87\, (2.93)$ GeV$^2$/fm for $c = 1$ and
$T = 200\, (300)$ MeV. These values are just slightly smaller than
those in LRW. Yet, the 5d origin of (\ref{dcuatro}) is reflected in
the linear dependence in the `t Hooft coupling. In the following
sections we shall apply the expressions (\ref{gjetq}) and (\ref{formulita})
to another couple of relevant backgrounds.

Another interesting use of the renormalized expression (\ref{formulita}) is
the possibility to extend the analysis  to quark sources of finite mass.
Indeed, from the point of view of the derivation, there is nothing peculiar
about the integral upper limit being at the boundary, and it could equally
well extend up to a finite value $\rho_{m}  = r_m/r_H <\infty$. The physics
then is dual to a geometrical setting in which the ends of the fundamental
string
are attached  to a probe brane that is placed in the above background at a
fixed distance from the stack, $r_{m} = m\, \alpha'$, set up by the mass of
the quarks in the fundamental representation \cite{Karch:2002sh,Kruczenski:2003be}.
It is quite evident from the analytic form of the formula (\ref{formulita})
that cutting the value of the integral will decrease the denominator, and
hence enhance the value of $\hat q$. Popular scenarios include cases where
a probe D7 (alternatively a D6 or a D8) are placed inside D3 (respectively
D4) backgrounds.
\FIGURE[h]{\epsfig{file=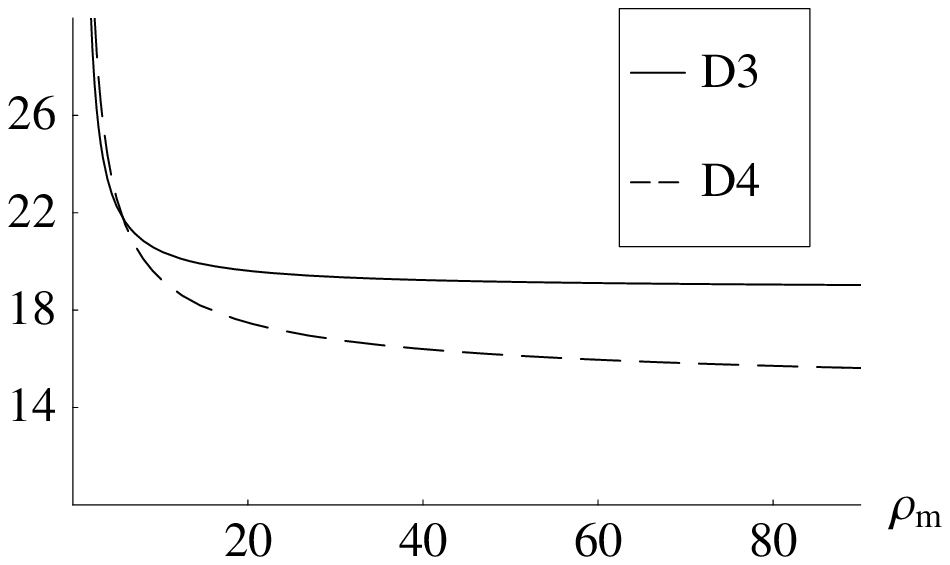,width=11cm}
\caption{Evolution of $\hat q$ with $m$ for finite mass quarks. The solid
curve represents the D3 background and the vertical axis is $\hat
q/(\sqrt{\alpha_{SYM}\,N_c}\,T^3)$ which asymptotes to $18.87\dots$ as in LRW.
The dashed curve represents the D4 background. The vertical axis is now
$\hat q/(c\,\alpha_{SYM}\,N_c\,T^3)$ which asymptotes to $14.26\dots$. The
horizontal axis is $\rho_m =  m\alpha'/r_H$. In each case, the quotient
$\alpha'/r_H$ can be read off from (\ref{solverh}).}
\label{fig4}}
Plotting $\hat q$ as a function of the  mass of the quarks, we
observe a very weak dependence until the mass is rather low, where the
approximations are questionable. For example, for
the case of the D3 background, the horizontal axis is $\rho_m =
m/(2\pi^{3/2}\sqrt{\alpha_{SYM}N_c }T)\simeq m/2.7$ GeV for
$\alpha_{SYM}=1/2, N_c=3$ and $T = 200$ MeV. In order to increase $\hat q$
by a 10$\%$ we must lower $\rho_m$ until a value of $\simeq 10$,
hence $m\simeq 27$ GeV. 


\section{$\hat q$ at finite coupling}

The AdS/CFT correspondence is a statement that goes beyond the classical
limit of string theory. In this limit, it maps classical solutions of
supergravity to quantum field theory vacua in the strong coupling limit,
$\lambda \to \infty$. Corrections in $\lambda^{-1}$ are in direct
correspondence with those in powers of $\alpha'$ in the string theory
side\footnote{Corrections in $\alpha'$ to timelike or spacelike Wilson
lines used to compute the $q\bar q$ potential have appeared in \cite{qqbar}.}.
In this paper we shall use the solution given in \cite{Gubser:1998nz,Pawelczyk:1998pb} corresponding to the $\alpha'$ corrected near extremal D3--brane. The relevant
pieces of information of this solution can be casted as follows
\beqa
\hat c_T^2(\rho) &=&  \rho^2 (1-\rho^{-4}) (1 + \gamma\, T(\rho) + ...)\, ,
\nonumber\\[1ex]
\hat c_X^2(\rho) &=& \rho^2   (1 + \gamma\, X(\rho) + ...)\, ,\nonumber\\[1ex]
\hat c_R^2(\rho) &=&  
   \rho^{-2} (1-\rho^{-4})^{-1}\, (1 + \gamma\, R(\rho) + ...) ~,
\nonumber
\eeqa
to first order in $\gamma =
\frac{\zeta(3)}{8} \,(\alpha'/R^2)^3\sim 0.15 \,\lambda^{-3/2}$. Here,
following (\ref{deltar}), we have already extracted $\Delta_R = R^2/r_H^2$
factors, with $R^2 = \sqrt{\lambda}\,\alpha'$. The intervening functions
read
\beqa
T(\rho) &=&  \left(-75 \rho^{-4} - \frac{1225}{16}\rho^{-8} + \frac{695}{16}\rho^{-12}
\right)\, ,
\nonumber \\
X(\rho)  &=&  \left(-\frac{25}{16} \rho^{-8}(1 + \rho^{-4})\right)\, ,
\label{TXR}\\
R(\rho)  &=&  
 \left(75 \rho^{-4} + \frac{1175}{16} \rho^{-8} - \frac{4585}{16}\rho^{-12}
\right)\, .
\nonumber
\eeqa
Inserting this data in  (\ref{formulita}) with $p=3$ and expanding to first
order in $\gamma$ gives
\be 
\hat q(\lambda) = \hat q(0)
 \left(1 - \gamma \left[ \frac{I}{2a}+45 \right]+... \right)
\ee 
with $\hat q(0)$ as given in \cite{Liu:2006ug}, and
\beqa
a & = & \int_1^\infty\frac{d\rho}{\sqrt{\rho^4 - 1}} = \sqrt{\pi}\,
\frac{\Gamma\left(5/4\right)}{\Gamma\left(3/4\right)} ~, \\
I & = & \int_1^\infty\frac{R(\rho) - X(\rho)\,(\rho^4 + 2) +
T(\rho)\,(\rho^4-1)}{\sqrt{\rho^4-1}}d\rho = - \frac{30725\,
\pi^{3/2}}{924\sqrt{2}\,\Gamma\left(3/4\right)^2} ~.
\eeqa
Evaluating yields
\be
\hat q(\lambda)  = \hat q(0)
 \left(1 - 1.7652\, \lambda^{-3/2} + \dots \right).
\ee 
Therefore, we see that  finite coupling corrections tend to diminish
the value of the quenching parameter. For example, taking $N_c=3$ and
$\alpha_{SYM}= 1/4$, $1/2$ and $1$, we get a reduction factor of $6,$
$2$ and $0.8$ percent respectively. Note that the decrease in the jet
quenching parameter towards weak coupling is suggestive of a smooth
interpolation between the strong coupling regime and the perturbative
results. Obviously, the computation of higher order corrections would
be necessary to put this conclusion on more solid grounds.


\section{$\hat q$ at finite chemical potential}

The  near horizon metric of a rotating black D3--brane with maximal number
of angular momenta reads as follows \cite{Russo:1998by,Cvetic:1999xp}, with
the conventions of \cite{Buchel:2006dg}:
\be
ds^2 = \sqrt{\Delta}\left( - \HHH^{-1}  f dt^2 + 
f^{-1}dr^2  + \frac{r^2}{R^2}\, d\vec x\cdot d\vec x \right)
+ \frac{1}{\sqrt{ \Delta}} \sum_{i=1}^3 R^2 H_i \left[
d\nu_i^2 + \nu_i^2(d\phi_i
+ A^i dt)^2 \right] ~,
\label{rotmet}
\ee
where $\nu_1 = \cos\theta_1,~ \nu_2 = \sin\theta_1\cos\theta_2,~ \nu_3 =
\sin\theta_1\sin\theta_2$, and $\HHH = H_1 H_2 H_3$, where
\be
A^i = \frac{1}{R}\sqrt{\frac{\mu}{q_i}}(1-H_i^{-1}) ~, \qquad
H_i  = 1 + \frac{q_i}{r^2} ~, \qquad
f = \frac{r^2}{R^2}\HHH -\frac{\mu}{r^2} ~, \qquad
\Delta = \HHH\sum_{i=1}^3\frac{\nu_i^2}{H_i} ~.
\ee
Upon Kaluza-Klein reduction, this becomes a charged AdS black hole solution
of $\mathcal{N}=2$~$U(1)_R^3$ supergravity, where $A^i$ plays the r\^ole of
the gauge field. The holographic gauge theory corresponding to (\ref{rotmet})
is $\mathcal{N}=4$~$SU(N)$ supersymmetric Yang-Mills at finite temperature
and with a chemical potential for the $U(1)_R^3$ symmetry.
It will be convenient to trade the nonextremality parameter  $\mu$ for the
horizon radius, $r=r_H$ given as the largest root of $f(r_H)=0$, i.e.,
\be
\mu =  \frac{r_H^4}{R^2 }\, \HHH (r_H)   \, ,
\ee
and define the adimensional quantities
\be
\kappa_i = \frac{q_i}{r_H^2} ~, \qquad
\Delta_R =\frac{R^2}{r_H^2} ~.
\ee
As usual, we go to dimensionless variable $\rho = r/r_H$ and find
\be
H_i(\rho) = 1+ \kappa_i\rho^{-2} ~, \qquad
f(\rho) =   \frac{1}{\Delta_R}\left( \rho^2\HHH(\rho)-\rho^{-2}\HHH(1)\right)
\equiv \frac{1}{\Delta_R}\hat f(\rho) ~.
\ee
Finally, the relevant functions entering the formula (\ref{gjetq}) can be
easily extracted from (\ref{rotmet}):
\be
\hat c_T^2(\rho) = \frac{ \sqrt{\Delta}\,\hat f}{\HHH} -
\frac{1}{\sqrt\Delta}\sum_{i=1}^3 \frac{\nu_i^2\, \HHH(1)}{ \kappa_i
H_i}(H_i-1)^2 ~, \qquad \hat c_X^2(\rho) = \sqrt{\Delta}\,\rho^2 ~,
\qquad \hat c_R^2(\rho) = \frac{\sqrt{\Delta}}{\hat f} ~.
\ee
The factors in the metric depend on the internal angles. However, the terms
above conspire to give
\be
\int_{1}^\infty\frac{\hat c_R d\rho}{\hat c_X^2\sqrt{\hat c_X^2 -
\hat c_T^2} } = \frac{1}{\HHH(1)} \int_{1}^\infty d\rho \left(
\rho^4\frac{\HHH(\rho)}{\HHH(1)}-1 \right)^{-1/2} ~,
\ee
where all information about the internal angular coordinates has dissapeared.
Now, given that the Hawking temperature of this solution
is given by \cite{Russo:1998by}
\be
T  = \frac{2 + \sum_{i=1}^3 \kappa_i - \prod_{i=1}^3\kappa_i}{2\sqrt{\HHH(1)}} ~\frac{r_H}{\pi R^2}
\label{Temp}
\ee
with $R^2 = \sqrt{\lambda}\,\alpha'$, substituting in (\ref{gjetq}) we find
the answer
\be
\hat q(\kappa_i) 
= \frac{\pi^2   T^3 \sqrt{\lambda}}{\sqrt{2}}\, \HHH(1)\,
 \left(\frac{2\sqrt{\HHH(1)}}{2 + \sum_{i=1}^3 \kappa_i - \prod_{i=1}^3\kappa_i}\right)^3
 \left(
 \int_{1}^\infty d\rho \left( \rho^4\frac{\HHH(\rho)}{\HHH(1)}-1 \right)^{-1/2}
 \right)^{-1}.  \label{qkappa}
\ee
In order to analyze this result, it must be recalled that the domain of thermodynamical stability of this solution \cite{Cvetic:1999ne,Son:2006em} (see also
\cite{Cai:1998ji}) is bounded by the inequality\footnote{A wrong sign
in this expression in a previous version of this paper has led to wrong
plots that we have corrected in this version.} $\kappa_1 + \kappa_2 + \kappa_3
- \kappa_1\, \kappa_2\, \kappa_3 < 2$.
\FIGURE[h]{\epsfig{file=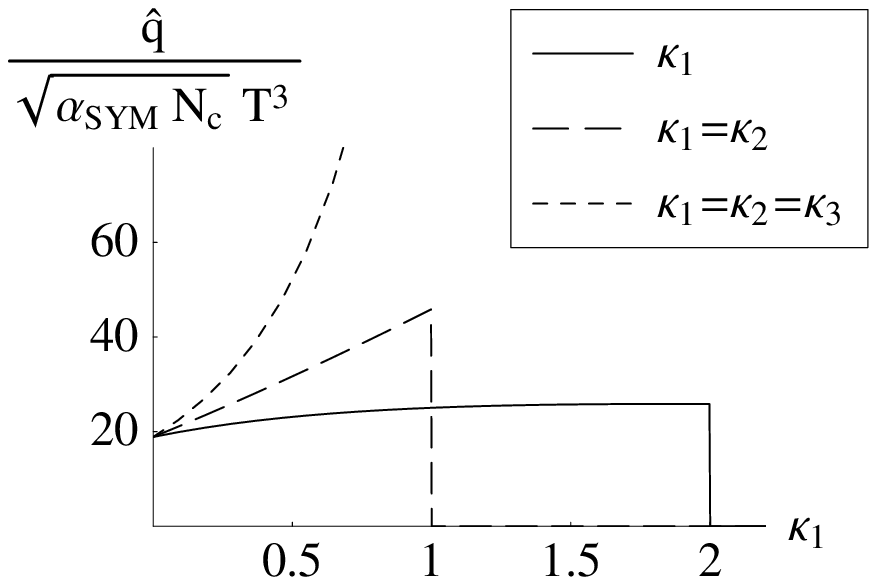,width=11cm}
\caption{Plot of $\hat q$ as  function of different combinations of charges
$\kappa_i$. The charges which are not varied are set equal to zero. The
stability bound chops the lower curves, but leaves the upper
one unbounded. At the origin the value is $\simeq 18.87$ as found in LRW.}
\label{fig1}}
Plotting the right hand side of (\ref{qkappa}) numerically, we find the curves that
are shown in figure \ref{fig1}.
The jet quenching parameter raises its value for nonzero charges. The increase is not monotonous along the whole space of charges. In fact, though hardly noticeable, it changes the sign of the slope along the line $\kappa_1$ with $\kappa_2=\kappa_3=0$. To see this better, we have zoomed the interval $\kappa_1\in [1.4,2]$ in figure \ref{fig2}, where we can see that the change in slope sign happens around
$\kappa_1 = 1.8$.
\FIGURE[h]{\epsfig{file=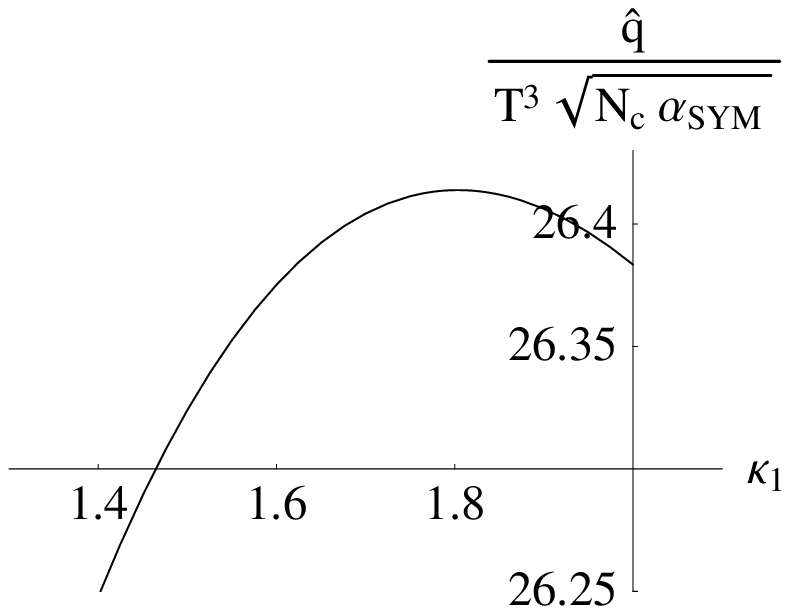,width=11cm}
\caption{Zoom of the curve $\hat q(\kappa_1)$ showing the value of the
turning and end points.}
\label{fig2}}

A comparison with recent results in the literature is in order. We certainly
agree and go beyond the perturbative analysis of C\'aceres and Guijosa
\cite{Caceres:2006as}. Also, when restricted to the cases examined by Lin
and Matsuo \cite{Lin:2006au} (one charge) and Avramis and Sfetsos
\cite{Avramis:2006ip} (one or two equal charges), we find qualitative
agreement within the range of stability. We may also examine less
symmetrical configurations on a 3 dimensional plot in figure \ref{fig3}.
\FIGURE[h]{\epsfig{file=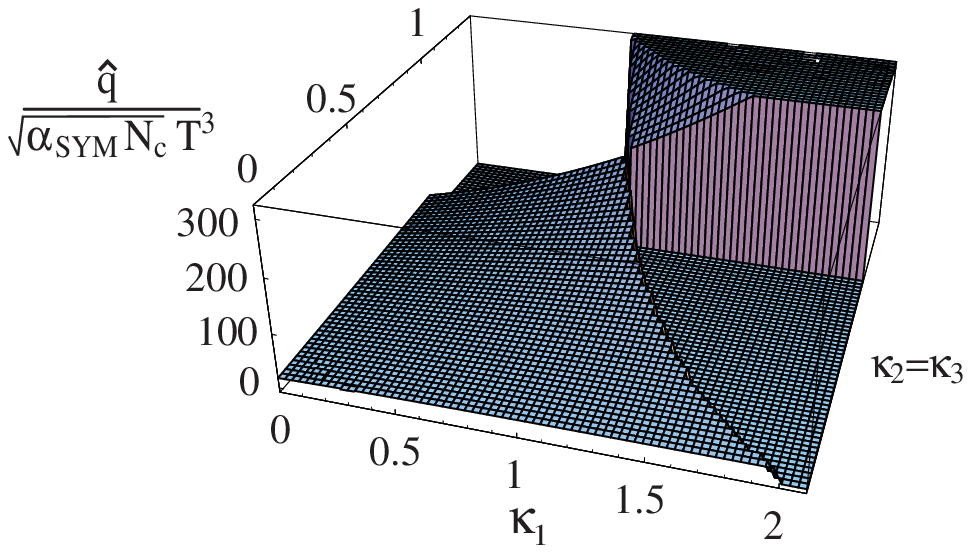,width=12cm}
\caption{Evolution of $\hat q$ in a sector of the plane $(\kappa_1,
\kappa_2 = \kappa_3)$. The curve is chopped by the thermodynamical stability
bound.}
\label{fig3}}

Besides analyzing the qualitative behaviour of $\hat q(\kappa_i)$ numerically,
in order to compare with other approaches, it may be of interest to
perform an expansion in powers of quantum field theoretical magnitudes.
For this purpose, it is relevant to recall how the thermodynamical magnitudes
are related to geometrical quantities. In particular, the density of physical
charge and chemical potential are given respectively by \cite{Son:2006em,Mas:2006dy}
\beqa
\rho_i &=& \frac{\pi N^2 T_0^3}{8}\sqrt{2\kappa_i}\, \prod_{i=1}^3(1 +
\kappa_i)^{1/2} ~, \\
\mu_i &\equiv& \left. A^i(r)\right\vert_{r=r_H} = \frac{\pi
T_0\sqrt{2\kappa_i}}{1 + \kappa_i} \prod_{i=1}^3(1 + \kappa_i)^{1/2} ~,
\eeqa
where $T_0 = r_H/(\pi R^2)$. From these expressions and (\ref{Temp}), we
should invert $\kappa_i$ in terms of $\rho_i$ and $T$ for the canonical
ensemble and in terms of $\mu_i$ and $T$ for the grand canonical ensemble.
This is dificult in the general case, so we may simplify for equal or
vanishing values of $\kappa_i$. For example, taking  $\kappa_1=\kappa$
and $\kappa_2=\kappa_3 = 0$  the two (inverse) expansions yield
\beqa
\kappa_C &=& \xi  - \xi^2 +\frac{11}{4}\xi^3 +...~,  \\
\kappa_{GC} &=& \zeta+\zeta^2 + \frac{5}{4} \zeta^3+... ~,  
\eeqa
with 
\be
\xi = \left(\frac{4\sqrt{2}\rho}{\pi N^2 T^3}\right)^2 ~, ~~~~~~~~~~~
\zeta = \left(\frac{\mu}{\sqrt{2}\pi T}\right)^2 ~. 
\ee
Expanding (\ref{qkappa}) in powers of $\kappa$ and inserting these series we obtain for the canonical and grand-canonical ensemble respectively the
following results
\beqa
\hat q_{C}(\rho) &=& \hat q (0)\left(1 + 0.63\, \xi - 1.08\, \xi^2 + 2.83 \xi^3 + ...\right) ~, \\
\hat q_{GC}(\mu) &=& \hat q(0)\left(1 + 0.63\, \zeta +0.18\, \zeta^2 + 0.06 \zeta^3 + ... \right) ~.
\eeqa
This expansion fully agrees with the one in \cite{Avramis:2006ip} upon
rescaling\footnote{The rescaling by 2 in $\zeta$ is to be traced to the
normalization of the gauge fields in \cite{Son:2006em}, which is $\sqrt{2}$
larger than usual. We thank Spyros D. Avramis for pointing this and some
calculational errors in a previous version.} $\xi\to 2\xi^2$ and
$\zeta\to 2\hat\xi^2$. 

~

The computation of Wilson loops in thermal $AdS$ backgrounds is a promising
line of research, of which the $\hat q$ computation is a salient example.
The identification of other observables within the AdS/CFT framework (see,
for a recent example, \cite{Peeters:2006iu}) that can be confronted with experimental data is an urget challenge.
It seems clear to us that there are several avenues for further exploration.
Among these, the extension of our results to less supersymmetric backgrounds
is of neat interest. We hope to report on these issues in the near future.

\medskip
\section*{Acknowledgments}
\medskip
We thank Alex Kovner, Krishna Rajagopal, Carlos Salgado and Urs Wiedemann
for useful discussions. Special thanks go to Alfonso V. Ramallo who was involved
in the first stages of this work, for his insightfull comments and support.
NA was supported by Ministerio de Educaci\'on
y Ciencia of Spain under a contract Ram\'on y Cajal, and by
CICYT of Spain under project FPA2005-01963.
JDE and JM were supported in part by MCyT, FEDER and Xunta de Galicia under
grant FPA2005-00188 and by the EC Commission under grants HPRN-CT-2002-00325
and MRTN-CT-2004-005104. JDE was also supported by the FCT grant
POCTI/FNU/38004/2001 and by Ministerio de Educaci\'on y Ciencia of Spain
under a contract Ram\'on y Cajal.  
Institutional support to the Centro de Estudios Cient\'\i ficos (CECS) from
Empresas CMPC is gratefully acknowledged. CECS is a Millennium Science
Institute and is funded in part by grants from Fundaci\'on Andes and the
Tinker Foundation.


\end{document}